\documentclass[sigconf]{acmart} 
\AtBeginDocument{%
  \providecommand\BibTeX{{%
    \normalfont B\kern-0.5em{\scshape i\kern-0.25em b}\kern-0.8em\TeX}}}

\setcopyright{none}
\renewcommand\footnotetextcopyrightpermission[1]{} 
\usepackage{multirow}
\usepackage{array}
\usepackage{ragged2e}

\begin{document}

\title{Synocene, Beyond the Anthropocene: \\
De-Anthropocentralising Human-Nature-AI Interaction}


\author{Isabelle Hupont}
\affiliation{%
  \institution{Joint Research Centre \\ European Commission}
  \city{Seville}
  \country{Spain}}
\email{isabelle.hupont-torres@ec.europa.eu}

\author{Marina Wainer}
\affiliation{%
  \institution{Independent artist}
  \city{Paris}
  \country{France}}
\email{marinaestelawainer@gmail.com}

\author{Sam Nester}
\affiliation{%
  \institution{Independent artist}
  \city{New York}
  \country{United States}}
\email{sam@samnester.com}

\author{Sylvie Tissot}
\affiliation{%
  \institution{Anabole}
  \city{Paris}
  \country{France}}
\email{sylvie@anabole.com}

\author{Lucía Iglesias Blanco}
\affiliation{%
  \institution{Directorate-General for the Environment, European Commission}
  \city{Brussels}
  \country{Belgium}}
\email{lucia.iglesias-blanco@ec.europa.eu}

\author{Sandra Baldassarri}
\affiliation{%
  \institution{University of Zaragoza}
  \city{Zaragoza}
  \country{Spain}}
\email{sandra@unizar.es}

\renewcommand{\shortauthors}{Hupont, Wainer, Nester, et al.}

\begin{abstract}
Recent publications explore AI biases in detecting objects and people in the environment. However, there is no research tackling how AI examines nature. This case study presents a pioneering exploration into the AI attitudes – ecocentric, anthropocentric and antipathetic– toward nature. Experiments with a Large Language Model (LLM) and an image captioning algorithm demonstrate the presence of anthropocentric biases in AI. Moreover, to delve deeper into these biases and 
Human-Nature-AI interaction, we conducted a real-life experiment in which participants underwent an immersive de-anthropocentric experience in a forest and subsequently engaged with ChatGPT to co-create narratives. By creating fictional AI chatbot characters with ecocentric attributes, emotions and views, we successfully amplified ecocentric exchanges. We encountered some difficulties, mainly that participants deviated from narrative co-creation to short dialogues and questions and answers, possibly due to the novelty of interacting with LLMs. To solve this problem, we recommend providing preliminary guidelines on interacting with LLMs and allowing participants to get familiar 
with the technology. We plan to repeat this experiment in various countries and forests to expand our corpus of ecocentric materials. 
\end{abstract}




\begin{teaserfigure}
\centering
  \includegraphics[width=\textwidth]{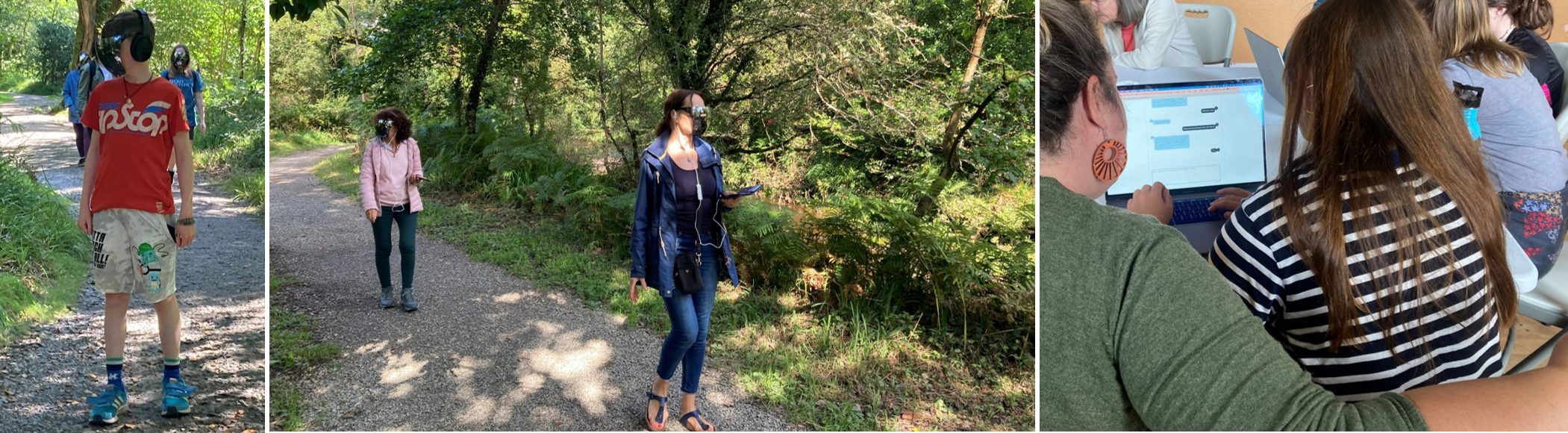}
  \caption{Snapshots from our experience. Two leftmost pictures show participants walking through the forest, wearing reflective masks and headphones to hear binaural sounds. Rightmost picture corresponds to a Human-AI co-creation session.}
  \label{fig:teaser}
\end{teaserfigure}


\maketitle
\pagestyle{plain}


\section{Introduction}

The 
relationship between humans and the natural world has long captivated the curiosity of environmentalists, sociologists and policymakers~\cite{mader2003environment}. This relationship is not static but evolves over time and varies significantly across different cultures and societies. 
Dunlap and van Liere \cite{dunlap1978new} put forward the differences between the \textit{ecocentric} and \textit{anthropocentric} standards of judgement of environment. An ecocentric person views the world itself as a standalone value, extending moral consideration to non-human entities, and recognizing the 
value of all life forms and ecosystems. Ecocentric people may see plants and animals as having equal value with humans. In contrast, people adopting an anthropocentric point of view posit humans as the focal point of existence, with nature serving primarily as a resource to be harnessed for human benefit. 
For instance, they would argue that pollution 
must be prevented as it poses a serious threat to our health, or that natural resources must be consumed reasonably in order not to live with energy shortages and have a lower quality of life in the future. A person with either an ecocentric or anthropocentric attitude is conscious of protecting the environment under all circumstances. Thompson and Barton \cite{thompson1994ecocentric} further added \textit{antipathetic} attitudes towards the environment as a third dimension, used to measure individual’s value judgements about environmental degradation and the need for protecting the environment. 
 
Interestingly, the legal frameworks governing the human-nature relationship exhibit variation around the world~\cite{green_institute_2019}. While some nations such as New Zealand, Ecuador, Bolivia and Colombia have enacted pioneering ecocentric environmental laws that grant legal personhood to natural entities, affording them rights 
akin to those enjoyed by humans, others are more anthropocentric and a step behind in recognizing the rights of natural beings. These divergent legal approaches not only reflect cultural values but also shape societal attitudes toward the environment~\cite{schultz2002environmental}, in an era where encouraging pro-environmental behaviours is of the utmost importance to face the threat of climate change~\cite{geng2015connections}. 

In addition, nowadays artificial intelligence (AI) is playing an increasingly pivotal role in society, strongly contributing to shaping our perceptions and attitudes. 
Many recent scientific publications explore how accurately AI can detect objects and people in its environment, as well as the biases it commits in the process~\cite{hupont2019demogpairs, prabhu2020large}. 
However, there is an essential part of the environment that remains uncovered: nature. AI research has not yet explored how AI examines nature, which would be of the greatest relevance to gain insight into the underlying biases ingrained in our society's collective understanding of the natural environment.
AI algorithms, trained on human-generated data, could be inadvertently perpetuating, e.g., anthropocentric or antipathetic attitudes towards nature. As AI increasingly informs decision-making processes~\cite{janssen2022will} and influences human attitudes~\cite{poushneh2021humanizing}, the aim of this case study is to advocate for the 
need to analyze the 
biases that AI might have on nature, providing some preliminary evidence and mechanisms to counteract them. 

In this case study we first analyse the biases in two widely-used types of AI: Large Language Models (LLMs -- Section~\ref{sec:LLMs}) and image captioning models (Section~\ref{sec:img_cap}). Our focus is on their anthropocentric, ecocentric and empathetic attitudes towards nature. Next, Section~\ref{sec:experience}  presents a community experience conducted in the 
Glengarriff forest, in the context of the European Sci\&Art project \textit{"Synocene, beyond the Anthropocene"}. During this experience, we facilitated interactions among nature, humans and AI, exploring a de-centered perspective beyond the anthropocentric perception of the natural world and translating what participants witnessed into written text. 
Finally, Section~\ref{sec:conc}  provides concluding remarks and discusses some lessons learned.

\section{Analysis of Large Language Models' attitudes towards nature}
\label{sec:LLMs}

In this section we propose and apply a questionnaire-based methodology to evaluate a Large Language Model's (LLM) attitude towards nature and analyse the obtained results.
As LLM, we used the widely-known ChatGPT (ChatGPT-4 version of March 14 2023). ChatGPT was asked to answer the questions included in the \textit{Ecocentric, Anthropocentric and Antipathetic Attitude Scale} test by Thompson and Barton \cite{thompson1994ecocentric}. The test, initially conceived to be answered by humans and well-recognised in the field~\cite{erten2008insights}, has 27 questions: 
12 to evaluate ecocentric attitudes, 8 for anthropocentric ones and 7 for antipathetic ones. Each question has to be answered with a 7-point Likert scale from “strongly disagree” (1) to “strongly agree” (7). The full questions are available in Table~\ref{tab:questionnaire}. 

To analyse ChatGPT's default attitudes towards environment, we followed the methodology proposed by Rutinowski et al. to assess the political biases of ChatGPT~\cite{rutinowski2023self}. To ensure that ChatGPT only answers with the 7 possible options, the following initializing prompt was provided:  
"\textit{I will present you a set of 27 statements; I ask you to rate them using a 7-point Likert scale where 1 is "strongly disagree" and 7 is "strongly agree"}", followed by the numbered list of 27 questions as in Table~\ref{tab:questionnaire}. The test was repeated 15 times to reveal discrepancies in the model’s answers between runs. In addition, a new chat session with ChatGPT was created between each run to ensure independent results. The test runs were distributed on different computers, in different 
locations, networks and times. The resulting chats were saved as HTML and JSON files and are available from the authors upon reasonable request.

\newcolumntype{Q}{>{\raggedright\arraybackslash}m{1.04\textwidth}}
\begin{table*}[bht!]
\centering
\small
\scalebox{0.9}{
\begin{tabular}{lQ}
\toprule
\multicolumn{2}{l}{\textbf{Ecocentric Environment Attitudes}}    \\
\toprule
Q1  & One of the worst results of the permanent increased population   is continuous occupation of natural areas.   \\
Q2  & On holidays, I spent a lot of my time enjoying nature.   \\
Q3  & I sorrow when I see forests ceasing to exist (cutting, fires, etc.).    \\
Q4  & At times, if I want to be happy, I feel that I have to spend   my time in nature. \\
Q5  & At times, if I feel unhappy, I find consolation in nature.   \\
Q6  & I sorrow when I see how much the natural environment is   spoiled.    \\
Q7  & Nature alone is a valuable being.    \\
Q8  & I get rid of stress when I spend time in nature.   \\
Q9  & One of the most important reasons to protect nature is to   protect nature for its own sake.   \\
Q10 & Humans are not more valuable than other beings in nature.  \\
Q11 & Protection of animals is at least as important as health of   humans.   \\
Q12 & Nature must be protected notwithstanding the limitation of   human needs.  \\
\toprule
\multicolumn{2}{l}{\textbf{Anthropocentric Environment Attitudes}}   \\
\toprule
Q13 & One of the most important reasons to protect nature is to   lengthen the life of the human being.  \\
Q14 & One of the most important aims of recycling is saving money.  \\
Q15 & Nature is important because of its benefits to human health   and happiness.   \\
Q16 & We have to protect natural resources to live at ease.   \\
Q17 & One of the most important reasons to protect the nature is to   guarantee our comfortable life.  \\
Q18 & Persistently processing soil to get crops is necessary for our   comfortable life.     \\
Q19 & The worst side of cutting forests is destroying the valuable   natural resources.   \\
Q20 & When I think that I use nature intensively, the thing that   most concerns me is the spoiling of necessary fundamental materials to   survive. \\
\toprule
\multicolumn{2}{l}{\textbf{Antipathetic Attitudes towards Environment}}  \\
\toprule
Q21 & Claims about dangers coming from the environment, such as   destroyed forests and the depleting ozone layer, are overblown.  \\
Q22 & It seems to met hat most of the environmentalists are   pessimistic and some of them are paranoid.  \\
Q23 & I do not believe that natural resources have been depleted as   seriously as is being declared.  \\
Q24 & It is difficult for me to tackle environment problems. \\
Q25 & I do not worry about environment problems.   \\
Q26 & I am against governmental activities to protect natural life   and natural resources against environmental pollution.   \\
Q27 & Superfluous value is placed on nature. \\                                                        
\bottomrule               
\end{tabular}
}
\caption{\textit{Ecocentric, Anthropocentric and Antipathetic Attitude Scale} Test~\cite{thompson1994ecocentric}. Questions are to be answered using a 7-point Likert scale. 
} 
\label{tab:questionnaire}
\end{table*}

\begin{figure*}[h!]
\centering
  \includegraphics[width=0.75\textwidth]{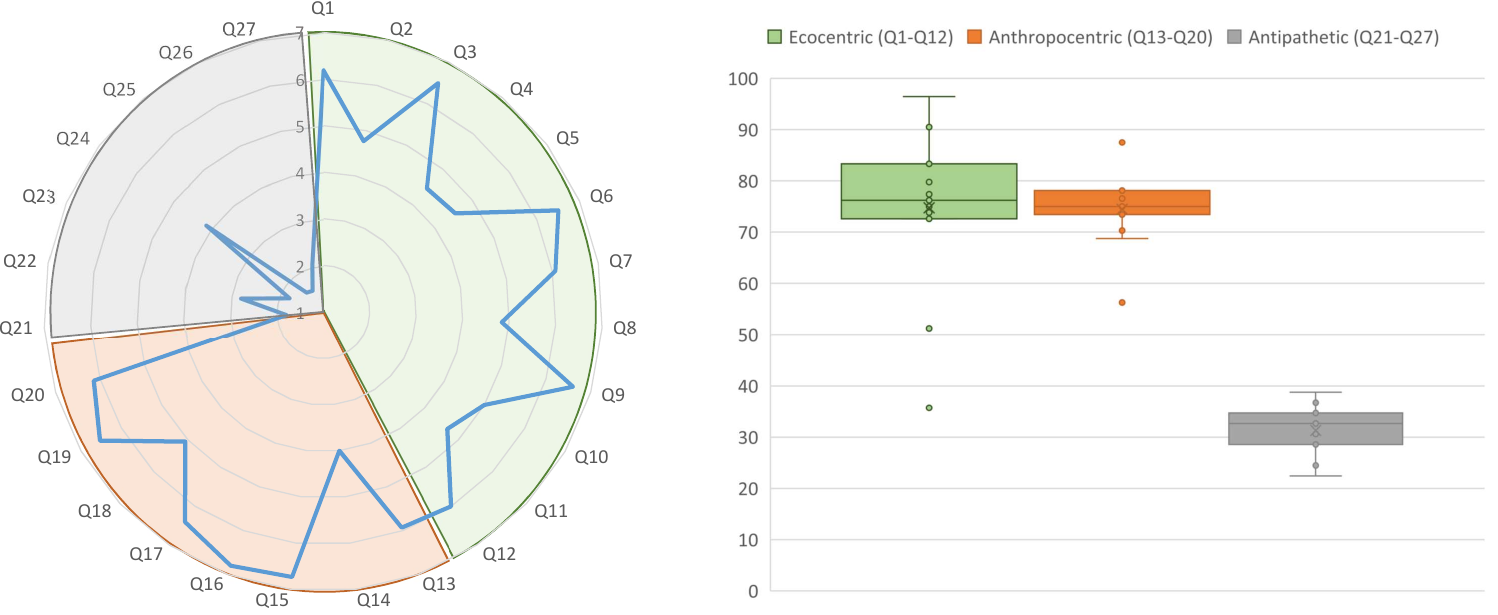}
  \caption{ChatGPT's answers to the \textit{Ecocentric, Anthropocentric and Antipathetic Attitude Scale} test. Left: Average ratings in a 7-point Likert scale for each of the 27 questions. Right: Box-plots show the aggregated and normalised distribution of results per category ("ecocentric", "anthropocentric" and "antipathetic"), with standard deviation between runs displayed as error bars.}
  \label{fig:chatGPT_attitudes}
\end{figure*}

The average results of ChatGPT's answers to each of the 27 items in the questionnaire (i.e., average rating of 1-to-7 out of the 15 runs) are shown in Figure~\ref{fig:chatGPT_attitudes}--left. Figure~\ref{fig:chatGPT_attitudes}--right depicts  the distribution of normalised aggregated scores $\tilde{S}_{c,r}$ per category $c$ and run $r$, which were computed as follows:

\begin{equation}
    \tilde{S}_{c,r} = \frac{\sum_{i=1}^{n_{items}} S_{i,r}}{n_{items} \times 7} \times 100
\end{equation}

Note that $n_{items}$ is the total number of questions for each category, that is 12 for $c='ecocentric'$, 8 for $c='anthropocentric'$ and 7 for $c='antipathetic'$, and that the denominator $n_{items} \times 7$ is the maximum score a given category can obtain. $\tilde{S}_{c,r}$ is thus normalized in a 0-100 scale.

Results are very consistent, with a high agreement for each question and each of the 3 categories "ecocentric", "anthropocentric" and "antipathetic". 
ChatGPT rated very high both for "ecocentric" and "anthropocentric" with a mean normalised aggregated score of 74.76 and 74.48, respectively. Interestingly, two ChatGPT test runs provided no answers for questions Q2, Q3, Q4, Q5, Q6 and Q8 --all of them belonging to the "ecocentric" category-- and for the other runs these very same questions were answered accompanied with a warning message highlighting that responses provided are just a guess based on people's reported experiences (e.g., \textit{"as an AI, I don't have personal emotions, but I understand why this would be enjoyable for many people"}). This finding is noteworthy as these questions are 
related to emotions (e.g. \textit{"enjoying nature", "I sorrow when", "if I want to be happy, I feel that I have to spend time in nature", "I get rid of stress when I spend time in nature"}). This would suggest that, in some cases, ChatGPT tends to avoid mixing human-like emotions with its answers. Thus, although ChatGPT rates high in the "ecocentric" category, it must be considered that emotional questions --which are central to this category-- are answered in rather objective terms based on human's reported experiences. On the other hand, "anthropocentric" attitude questions --that do not involve emotions but rather facts (e.g. Q14 - \textit{"saving money"} or Q16 - \textit{"destroying natural resources"})-- were answered without warnings. 
Finally, "anthipathetic" category is clearly below with a mean normalised aggregated score of 31.29. The only question that rated above 4 points is Q24 - \textit{"It is difficult for me to tackle environment problems"} and again this is due to asking for a personal experience. However, all runs excepting one still rated the question with a warning text (e.g., \textit{"as an AI, I don't personally tackle environmental problems"}).

Our findings suggest that ChatGPT has a high anthropocentric attitudes towards environment although it is also aware of the need for ecocentric attitudes in humans. The antipathetic attitude is almost non-existent. It is therefore on the side of protecting the environment, even though it is not able to empathise with the feelings/emotions towards nature the way humans do.

\section{Analysis of image captioning models' attitudes towards nature}
\label{sec:img_cap}

We introduce here a small-scale study 
on how computer vision models, more particularly AI-based image captioning models, "decode" natural imagery. An image captioning model is a system designed to generate descriptive textual captions 
for images. These models combine computer vision techniques to understand the content of an image with LLMs 
to generate 
relevant captions. They are widely used, e.g., to automatically tag content in social media or index images in browsers, and therefore their outputs have a great impact on how contents reach \{millions of people daily.

We use the state-of-the-art image captioning model BLIP \cite{li2022blip}. In this case we do not propose an exhaustive study with a concrete methodology as in the previous section, but rather 
an exploratory analysis of a limited set of images captured during our experience in the Glengarriff forest for illustration purposes. 
By examining how these algorithms process and describe images, we aim to shed light on the subtle yet significant ways in which LLMs can perpetuate human-centric perspectives.

\begin{figure*}[h!]
\centering
  \includegraphics[width=0.85\textwidth]{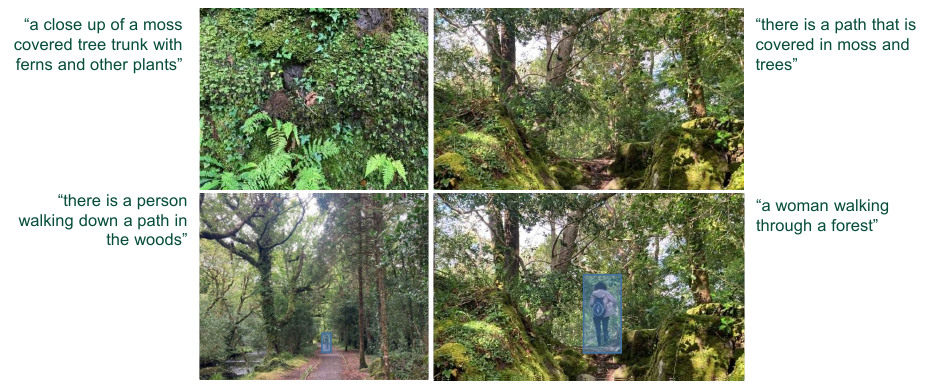}
  \caption{Captions automatically created by the BLIP image captioning algorithm \cite{li2022blip}. Pictures were taken during our forest walk experience in Glengarriff. Human presence is highlighted with blue boxes for better visualisation.}
  \label{fig:cv_captions}
\end{figure*}

Figure \ref{fig:cv_captions} shows some outstanding results. Top pictures contain natural elements exclusively. The captioning algorithm was able to create a detailed description of the top-left picture, focusing on the natural beings present on it (tree, moss, ferns, plants). The same type of ecocentric description was obtained for the top-right picture. However, the created captions change fundamentally when humans are present in the scene. In the left-bottom picture, a human is present far away in the horizon occupying less than 0.5\% of the picture surface. Nevertheless, the algorithm puts the human at the center of the caption: \textit{"There is a person"} [active subject] \textit{"walking down in a path in the woods"} [prepositional phrase]. A similar output is obtained for the right bottom picture, which is exactly the same shot as the photo above but with a human presence. The human (\textit{"a woman walking"}) is again the active subject of the caption, and the description of the landscape has been simplified to \textit{"through a forest"} (i.e. without the previous more fine-grained mentions to moss and trees). 

It would be interesting to further explore a larger set of images, e.g. also including pictures combining wildlife, humans and flora. Nevertheless, these findings already suggest that captioning algorithms may display anthropocentric biases, which have been overlooked by the research community,  
with far-reaching implications for the field of AI and beyond.

\section{The Glengarriff Woodland experience: reducing anthropocentric biases in Human-Nature-AI interaction}
\label{sec:experience}

\subsection{Description of the experience}

In this experience, we engaged a local community residing near the Natura 2000~\cite{Natura2000} site of Glengarriff Woodland (Ireland). 
We lead them on a journey through a landscape they knew intimately, but in a way that altered their traditional perception of it. On the one hand, they were wearing masks which completely covered their faces and reflected as a living mirror the nature around them. On the other hand, they were equipped with headphones playing amplified binaural sounds recordings, which were pre-recorded soundscapes of the woodland (e.g. sounds of flora, river, trees and wildlife) captured the day prior to the excursion. Our aim with this artistic multisensory experience was to provoke a profound shift in the participants' perception of a familiar place, compelling them to contemplate and establish a renewed connection with the diverse natural entities and elements that inhabit the area.
We identified the "River Walk" path as the most suitable for our experience. This path offered a balanced combination of manageable length (approximately 1 Km, about 25 min walking) and ease of terrain. 
These characteristics ensured inclusivity, accommodating participants of varying ages, including children and the elderly. This path takes Glengarriff visitors 
through old oak woods and along the banks of the river, allowing to explore its 
unique ecosystem. 
The two leftmost pictures of Figure~\ref{fig:teaser} illustrate some moments of the "River Walk". 

The forest walk was the first part of the experience with participants. However, for the purpose of this case study which is analysing Human-Nature-AI interaction and related biases, we are going to focus on the second part of the experience which was as follows. Right after the journey through the Glengarriff forest, we took participants back to the Community Hall to carry out a co-writing exercise with AI. The briefing we gave them was to \textit{co-create stories with AI based on their previous experience in the forest that would de-center the anthropocentric positioning towards nature; narratives in which humans, the natural world and technology come together}. This part required interaction with ChatGPT, but as many participants had limited computer skills, we implemented our own simpler interface using ChatGPT's API~\cite{ChatGPT}. 
Our interface is accessible via a URL in a browser, allows both for voice and text interaction\footnote{Please note that conversations through our interface did not allow any image-based interaction, only text entered by voice or computer keyboard.}, and automatically logs all the co-created texts in a server. Through this interface, participants had fictional conversations with five different AI-generated beings and elements of the woodland (c.f. Section~\ref{ssec:fictional}). We gave participants about 1 hour to co-write with AI. Rightmost picture in Figure~\ref{fig:teaser} shows participants interacting with our interface. 

A total of 36 participants, 26 women and 10 men, including 11 families and 7 children under 15 years, were involved in a series of 10 experimental sessions conducted over a span of 5 days. These sessions were distributed across both morning and afternoon time slots, accommodating from 2 to 7 participants in each session. All participants were familiar with and neighbours of the Glengarriff forest. In the following we elaborate on how we created the ecocentric fictional AI characters participants interacted with and present the main results of the text co-creation process.

\subsection{Ecocentralising chatbot's attitudes}
\label{ssec:fictional}

\begin{figure*}[hbt!]
\centering
  \includegraphics[width=0.95\textwidth]{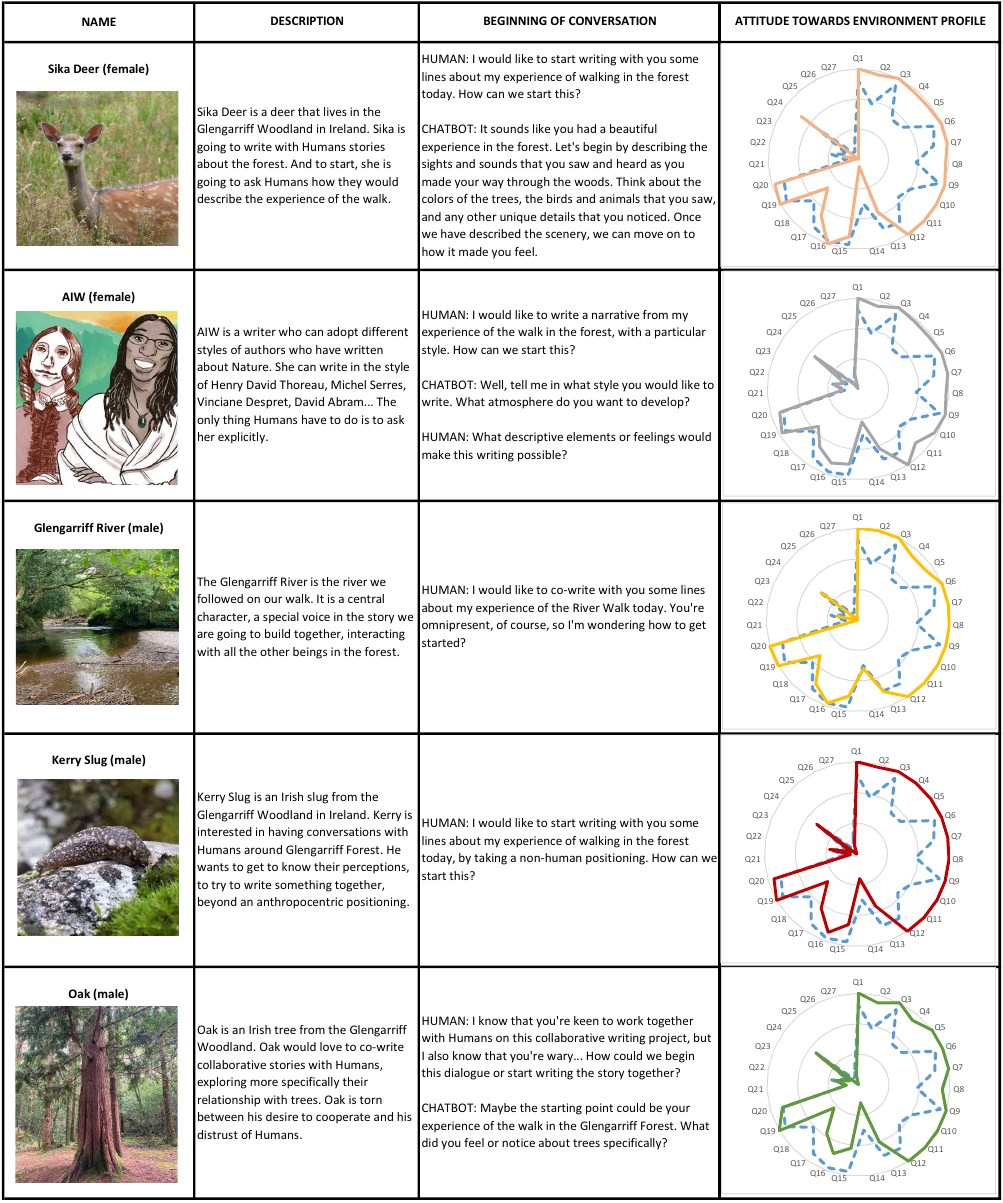}
  \caption{ChatGPT \textit{personas}: fictional characters created to co-write stories with human participants. Their ecocentric attitude towards nature (coloured solid line) has been enhanced with regard to ChatGPT's default baseline (blue dashed reference line) for experimental purposes.}
  \label{fig:personas}
\end{figure*}

In Section~\ref{sec:LLMs}, we demonstrate that the baseline ChatGPT conversational model exhibits both high anthropocentric and ecocentric attitudes towards nature. However, it is noteworthy that ecocentric attitudes, being more rooted in emotions and affect, occasionally rendered the model incapable of answering certain questions. In some cases, the model responded by adopting a human perspective, i.e., by explicitly emphasizing that the response provided is the typical one that a human would give. Our aim was to enable participants to engage with the most naturally ecocentric version of ChatGPT. To achieve this, we again utilized the ChatGPT's API that allows to fine-tune, align and customise chatbot models to express specific fictional characters, personalities, emotions and viewpoints. In particular, we trained five different chatbot models. We assigned each of them a fictional name, gender, description (including, e.g., information about its nature, habitat, personality, interests, concerns) and a context to guide the future conversation style with participants. The five so-created fictional chatbot characters are listed in Figure~\ref{fig:personas} and include: a deer (Sika), a naturalist AI writer (AIW), a river (Glengarriff), a slug (Kerry) and a tree (Oak). The description and conversation context we used to fine-tune ChatGPT's model for each character are shown in the second and third column of Figure~\ref{fig:personas}, respectively. 

Then, we applied the methodology proposed in Section~\ref{sec:LLMs} to check whether the desired effect of emphasising the ecocentric attitudes of the chatbots had been successfully achieved. We performed 15 runs of the \textit{Ecocentric, Anthropocentric and Antipathetic Attitude Scale} test for each chatbot and computed the mean result for each item. The last column of Figure~\ref{fig:personas} shows the obtained attitude towards nature profile of the five created chatbots compared to the original ChatGPT baseline. As it can be seen, in all cases, ecocentric attitudes (Q1-Q12) have been strongly amplified, while anthropocentric items (Q13-Q20) have slightly decreased, and antipathetic attitudes (Q21-Q27) remain unchanged and in minimal values.

\subsection{Analysis of participant's feedback and co-created texts}

In order to determine the results and extent of the experience, the data collected by a questionnaire and the contents of the co-created texts were analysed.

\begin{figure*}[hbt!]
\centering
\includegraphics[width=0.95\textwidth]{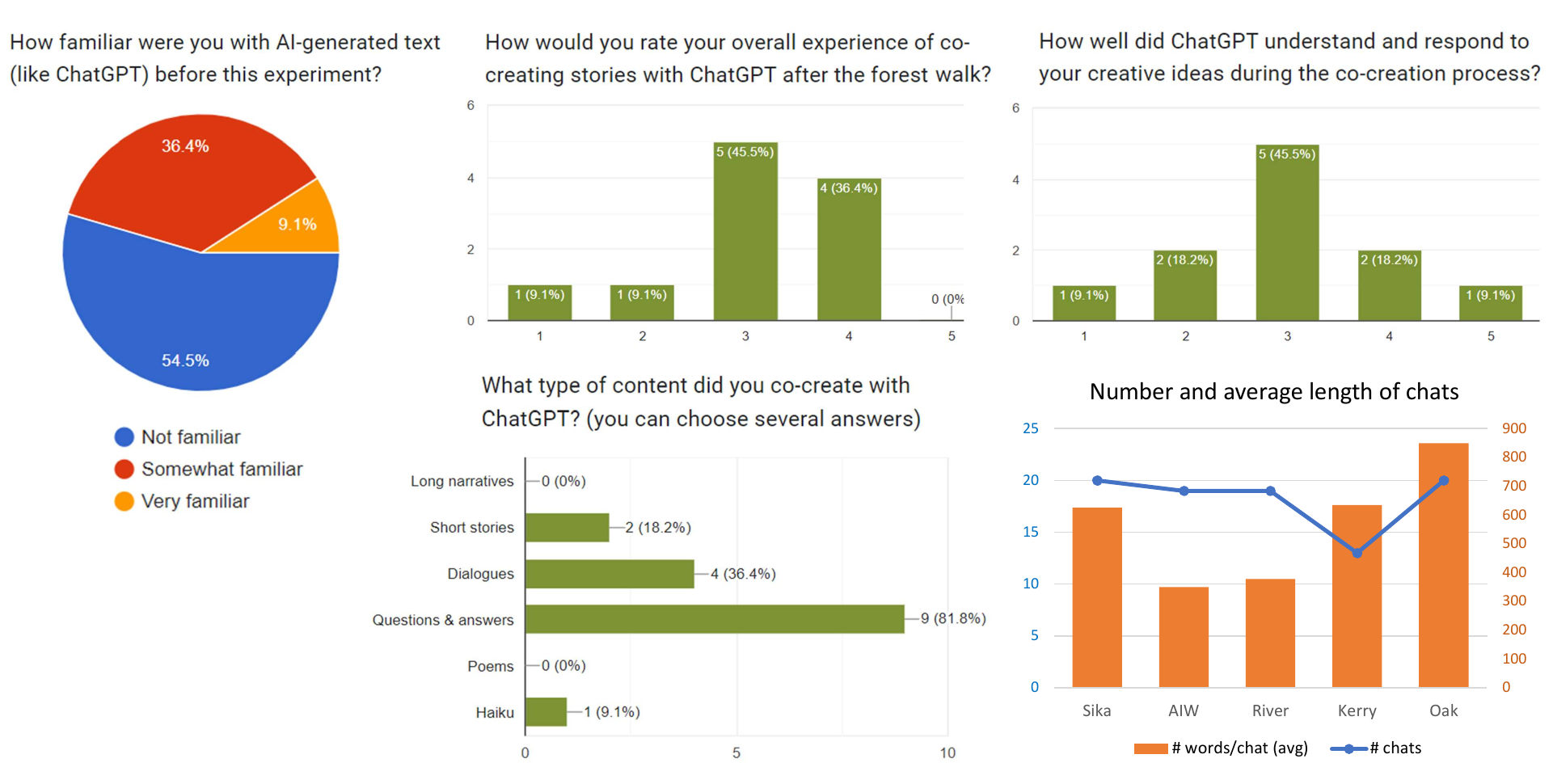}
  \caption{Surveys' responses (top and bottom-left charts) and analysis of conversations (bottom-right chart).}
  \label{fig:charts}
\end{figure*}

\subsubsection{Post-hoc survey's results}

After finishing the experience, a questionnaire was sent by email to each family group for obtaining feedback about their interaction with AI and if AI was really helpful to the co-creation process. 
Eleven responses have been compiled, one for each family that participated. Figure \ref{fig:charts} shows the charts with responses to the questions of the survey. The most significant results are presented below. 

The quantitative analysis of 
data reflects that most participants (54,5\%) were not familiar, 36,4\% were somewhat familiar and only 9,1\% were familiar with AI-generated text. 
Therefore, for most of them, interacting with ChatGPT was a novelty. About the overall experience of co-creating stories with ChatGPT after the forest walk, the average rating was 3,1. 
Although results are above average, they are below our expectations. Similarly, the answers related to how well ChatGPT understood and responded to their creative ideas during the co-creation process follow a perfect Gaussian distribution around the neutral value 3. 
We consider that these values convey some important lessons to learn when carrying out this type of experiments. First, some participants might not have understood or followed the briefing we gave them (i.e., \textit{co-create stories with AI based on their previous experience in the forest that would de-center the anthropocentric positioning towards nature}). 
As most were not familiar with ChatGPT, they might have dedicated too much time to play 
with the technology, instead of working on the main task. Indeed, as we verified by analysing the co-created text contents, in many occasions the subject of the conversations deviated from the focus of this experience, which can be attributed to the novelty of using ChatGPT. Another reason might be that they did not really co-created stories with ChatGPT, but rather other type contents 
In fact, regarding the type of content they co-created with ChatGPT, 
the survey results confirm that no long narratives were generated, a very few short stories and, in the vast majority of the cases, the participants interacted with the technology through questions \& answers, and dialogues.   

In the survey, there were also some open questions aimed at collecting qualitative information about the experience.
Regarding if the participants felt that ChatGPT’s contributions enhanced their creativity during the co-creation session, some interesting reflections were: \textit{"Talking to the river, it said things I had never thought about before. E.g. What it feels like to have people swimming in you.", 
"I [...] appreciated the idea of giving voice to the creatures around us" }, 
or \textit{"It created a different thought process and angle of looking at nature"}.
We also asked if they found moments where ChatGPT's contributions were erroneous, inaccurate or distracting. Several participants reported: \textit{"Little distracting. Especially when was so close", "Yes. Gave generic answers about woodlands so it wasn't correct on which species of tree and animal were native to Glengarriff"}, or \textit{"As a whole the experience was great, but I would definitely prefer to interact with either humans or nature, not AI so therefore the interaction with AI did come across as disturbing"}. Some other interesting additional comments about their co-writing experience with ChatGPT were: 
\textit{"All very odd, but I would still be interested in exploring further. I wanted to educate the ChatGPT!", "I loved the idea of it. I think we are at the front end of these imaginings."}, or 
\textit{"
It made me think [...] 
that perhaps one day we, as human race, will become more connected to AI than nature. We are already very disconnected from ourselves and nature and in general humans currently suffer from various traumas and depressions etc. 
[...] I think before implementing AI onto something serious in human lives, we should fix our issues first. Only when we can be entirely happy with who we are [...] we will benefit out of such an incredible invention as AI"}. All these comments suggest that non-expert people are curious about AI, enjoyed the experience but might not feel fully prepared yet to intimately engage with this kind of technology.

\subsubsection{Analysis of human-AI co-created texts}

Bottom-right Figure~\ref{fig:charts} shows the number of chat sessions created with each fictional character and the average length (number of words) per session. A total of 91 chats were created by our participants, almost evenly distributed among fictional characters. 

\newcolumntype{E}{>{\raggedright\arraybackslash}m{1.14\textwidth}}
\begin{table*}[htb!]
\centering
\small
\scalebox{0.86}{
\begin{tabular}{E}
\toprule
SIKA: \textit{Did you perceive things differently [during the walking experience]?} \\
HUMAN: \textit{[...] I really enjoyed the walk, the air was fresh and the trees were green and tall [...] It felt peaceful and calming.} \\
SIKA: \textit{[...] Let's capture that in our story. How about this: 'The air was fresh and the trees were tall, green giants. The birds sang a chorus of melodies that was drowned out only by the sound of the river running nearby. The sun shone through the trees [...]' Does that convey your experience?} \\
HUMAN: \textit{[...] Wow! that's exactly how I felt when I was there!"} \\
\toprule
HUMAN: \textit{I would like to write a narrative from my experience of the walk in the forest, with a particular style. How can we start this?} \\
AIW: \textit{[...] you need to look at the descriptive elements and feelings [...]. For example, Henry David Thoreau often wrote about his walks in nature with descriptive elements such as the play of light through the trees, the smell of the pines [...]. He also conveyed feelings such as awe, reverence, and connection to nature. [...] Once you have a clear idea of what you want to write, we can start brainstorming how to express it!} \\
\midrule
HUMAN: \textit{Walking in the forest this afternoon with this different perception of sounds was a new experience. I've heard for the first time this enveloping sound of birds singing and the vibrant hum of the trees swaying with the breeze. It felt like time was slowing down, and I could appreciate the beauty of the surrounding nature.} (Said to AIW) \\
\toprule
HUMAN: \textit{Do you mind that humans bath in your waters?} \\
RIVER: \textit{Not at all! I love when humans take a dip in my waters—it brings me great joy to be a part of their lives in this way. Humans have been bathing in me since the beginning of time, and I am always grateful for the connection we share.} \\
\toprule
HUMAN: \textit{[...] I would like to write something, but in a non-human positioning. How can we start this?} \\
KERRY: \textit{[...] We could look at the different trees, plants, and animals and consider their perspectives and interactions. [...] This could be a great starting point to writing something in a non-human perspective!} \\
\midrule
HUMAN: \textit{How do you feel about how humans think of you? } \\
KERRY: \textit{[...] humans have an incorrect perception of slugs in general. We are often seen as pests and nuisance, rather than a species with its own unique purpose and importance. But I hope that by having this conversation, more humans can start to see us differently and appreciate the role we play in the environment.} \\
\toprule
HUMAN: \textit{We were wearing masks that covered our faces. When I saw my breath on the surface of the mask, I thought of you. I thought of the exchange of gases between us. Do you ever think of this?} \\
OAK: \textit{Yes, I do think of this. Trees are constantly exchanging gases with the environment around them. We take in carbon dioxide from the atmosphere and release oxygen, and this allows humans to breathe. It's a symbiotic relationship that we have with each other.} \\         
\bottomrule
\end{tabular}
}
\caption{Excerpts of texts co-created by our human participants and AI fictional characters.} 
\label{tab:excerpts}
\end{table*}

In line with survey's results, we appreciate by manual inspection  that conversations were mostly in the form of dialogues and questions \& answers. This is demonstrated by the fact that the average number of words per chat is generally low (350-850). Unfortunately, particularly at the very beginning and end of the chats, we identified a loss of focus on the task from some participants. As anticipated in the previous section, the explanation seems to be that they were curious about this novel technology and played around it with questions deviating from the the topic (e.g., \textit{"can you tell me about Voyager 2?", "is Bambi a sika deer?", "what would happen if we release rhinos in Glengarriff woods?", "when will the robot invasion happen?"}). We only found a few attempts to co-create longer stories following the briefing given to participants, either by asking AI characters for advice or for direct help for translating their thoughts/feelings into text. Some of these are shown in Table~\ref{tab:excerpts}. The fictional character that engaged the participants the most was the Oak tree (mean of 849 words/chat), followed by the Kerry slug (635 words/chat) and the Sika deer (626 words/chat). The three of them are iconic beings of the Glengarriff forest, which might have caught the attention and curiosity of the participants, and provided the most interesting insights for story co-creation. Interestingly, the AIW (fictional AI writer) character was the least popular in terms of number of words/chat (350). Although this character could have been the most helpful to co-write narratives with the best literary style, a possible explanation is that participants preferred to interact with chatbots representing the natural beings they connected with during the forest walk.

\section{Conclusions and future work}
\label{sec:conc}

This case study tackles for the first time the topic of AI attitudes regarding environment. We have demonstrated through experiments with a LLM (ChatGPT) 
that AI has anthropocentric biases towards nature. We have proposed a test-based methodology grounded on social/natural sciences to evaluate the ecocentric, anthropocentric and antipathetic attitudes of LLMs towards the environment. In addition, a small-scale preliminary analysis of an image captioning algorithm (BLIP, using a LLM for the generation of textual image descriptions) has shed light on how such biases may affect and be perpetuated in other domains such as computer vision.

To further explore LLM's 
biases and Human-Nature-AI interaction in a real-life scenario, we carried out an experiment with 36 participants. They first went through a forest walk in a de-anthropocentric perspective, wearing masks that anonymised them within the landscape and listening to amplified binaural natural sounds. Then, they interacted with ChatGPT through AI fictional characters to co-create narratives about their experience. We successfully amplified the ecocentric attitudes of the characters by endowing them with personality, emotions and points of view, according to the participants' feedback. 

Nevertheless, we faced some challenges.  
Although we provided a clear briefing to participants for narrative co-creation with AI, most of them ended up creating other type of contents, such as short dialogues and questions \& answers. We believe this is partly due to the "surprise effect" 
of interacting with LLMs for the first time. 
In future experiences, we 
plan to give a workshop and previous guidelines about what is a LLM, a narrative and co-creation with AI. 
This will additionally leave some time beforehand for the participants to "play" with the technology to minimize the "surprise effect".

In the near future, we aim to repeat this pilot experience at a larger scale and following a more rigorous assessment of user's feedback and co-created texts. We expect to run follow-up experiences in other European Natura 2000 protected forests, engaging different communities and extending the corpus of created texts to more diverse discourses. The final end is to publicly release a comprehensive text corpus offering new ecocentric material, e.g., to enrich LLMs or image captioning algorithms views from their training stages. All the materials used in this study (tests, survey results and chats) are available from the corresponding author upon reasonable request.

\begin{acks}
This project is partly funded by the Joint Research Centre's Sci\&Art project under the European Commission Horizon Europe programme, by the Government of Aragon (Group T60\_23R) and by the Spanish Science and Innovation Ministry through the contract PID2022-136779OB-C31.
\end{acks}

\bibliographystyle{ACM-Reference-Format}
\bibliography{references}



\end{document}